\documentstyle[prl,aps,multicol,amsmath,amssymb,graphicx,psfrag]{revtex} 

\begin{document}

%%====================== DEBUT DU TEXTE ========================%% 
\newcommand{\fig}[2]{\includegraphics[width=#1]{#2}}
%%%%%%%%%%%%%%%%%%%%%%%%%%%%%%%%%%%%%%%%%%%%%%%%%%%%%
 \newcommand{\unecol}{
     \end{multicols}}
 \newcommand{\deuxcol}{
     \begin{multicols}{2}}
%%%%%%%%% MACROS Specifiques a l'article %%%%%%%%%%%%%
\newcommand{\psiL}[1] {\psi_{\text{\tiny L}#1}^{}}
\newcommand{\psiLd}[1]{\psi_{\text{\tiny L}#1}^{\dagger}}
\newcommand{\psiR}[1] {\psi_{\text{\tiny R}#1}^{}}
\newcommand{\psiRd}[1]{\psi_{\text{\tiny R}#1}^{\dagger}}
\newcommand{\psiLR}[1] {\psi_{\text{\tiny L/R}#1}^{}}
\newcommand{\psiLRd}[1]{\psi_{\text{\tiny L/R}#1}^{\dagger}} 
%%%%%%%%%%%%%%%%%%%%%%%%%%%%%%%%%%%%%%%%%%%%%%%%%%%%%%

\title{Two dimensional anisotropic non Fermi-liquid  phase of
coupled Luttinger liquids}
\author{Ashvin Vishwanath$^{\dagger}$ and David
Carpentier$^{\ddagger}$                                
} 
\address{
$^{\dagger}$ Physics Department, Princeton University, NJ 08544\\
$^{\ddagger}$ Institute for Theoretical Physics, University of
California, Santa Barbara, CA 93106--4030
}

\maketitle 

\begin{abstract}
 We show using bosonization techniques, that strong
forward scattering interactions between one dimensional spinless
Luttinger liquids (LL) can stabilize a phase where charge-density
wave, superconducting and transverse single particle hopping
perturbations are irrelevant. This new phase retains its LL like
properties in the directions of the chains, but with relations between
exponents modified by the transverse interactions, whereas, it is a
perfect insulator in the transverse direction.  The mechanism that
stabilizes this phase are strong transverse charge density wave
fluctuations at incommensurate wavevector, which frustrates crystal
formation by preventing lock-in of the in-chain density waves.
\end{abstract}

\deuxcol

Interacting fermions in one dimension can exhibit Luttinger liquid
behaviour where, in contrast to Fermi liquids, all the low lying
excitations are collective modes \cite{LL}.  The search for such
non-Fermi liquid behaviour in higher dimensions prompted several
authors to study the problem of coupled Luttinger liquids.  But
renormalization group (RG) studies, such as \cite{yakovenko92} found
the coupling to destabilize the Luttinger liquid behaviour. It was
however argued by Anderson {\it et al.} that, in spite of its relevance
in the RG sense, electron transverse hopping may be incoherent
\cite{anderson}, allowing for a non-Fermi liquid in dimensions greater
than one. In addition to these theoretical motivations, interacting 
Luttinger liquids could well arise in experimental systems like quasi
one dimensional organic conductors and in ``ropes'' of nanotubes. Recently,
 coupled one dimensional systems have also 
emerged in the stripe phases of Quantum Hall systems in higher Landau
levels\cite{QHEstripes}, and in the cuprates\cite{hitcstripes}.

 In this paper, we will revisit the problem of coupled Luttinger
liquids, and address the question of the existence of a stable two
dimensional phase that retains some of the properties of the one
dimensional Luttinger liquid. We consider this issue in detail within 
the RG framework using bosonization, for the case of spinless
fermions. Although most of the physical systems of interest are of
spinful (rather than spin polarized or spinless) fermions, that case
is technically more involved due to the presence of exchange
interactions between the chains, and we postpone its study to
\cite{inprep}. The case of an infinite set of coupled spinless
Luttinger liquids is simpler, but should still capture some of the
general physics present in the spinful case. There is also a
possibility that such systems may be directly realized by spin
polarizing a layered quasi-one-dimensional system of low electron
density with an in plane magnetic field; or in the spin polarized
Luttinger liquid formed by Zeeman split quasiparticles bound to the
superconductor vortex core \cite{avts,QHEstripefootnote}. The
new ingredient in our study is the inclusion of strong forward
scattering interactions between the chains. We then
find that an anisotropic phase, which is Luttinger liquid like along the
chains but a perfect charge insulator in the transverse direction,
 is stable against a range of instabilities including those that
could lead to a superconductor, crystal or 2D fermi liquid
phase. Hence this new phase is a non-Fermi liquid phase in two
dimensions that is highly anisotropic. In the direction of the chains
the correlation functions have power law forms with nontrivial
exponents, as in a Luttinger liquid. However, the relations between
exponents are modified, as compared with completely decoupled
Luttinger liquids, by the strong forward scattering interactions
between the chains. We will call this phase the {\it sliding Luttinger
liquid (SLL)}. It is analogous to the sliding phase found by 
O'Hern {\it et al.} \cite{ohern99} 
in the related problem of XY models coupled by suitable gradient
interactions, which motivated our approach. In contrast to
\cite{shulz}, 
we consider transverse hopping operators when establishing the
stability of the SLL.

The stable sliding Luttinger liquid fixed points that we find occur 
close to an instability towards transverse
charge density wave ordering. The instability occurs when the stiffness 
of the density fluctuation mode at a particular transverse wave-vector 
vanishes.  We propose a physical mechanism that is responsible for 
stabilizing this phase - i.e. that the proximity to the transverse
charge density wave state induces strong fluctuations of the local
in-chain density, which frustrates crystal formation. This mechanism is most 
effective if the wavelength of the transverse charge density wave phase is
incommensurate with the spacing between the chains.

The existence of this phase may at first sight contradict
previous results on coupled Luttinger liquids.
However, previous approaches focused  either on the simpler case
of two interacting chains \cite{yakovenko92}, or on the
perturbative effect of these transverse interactions. Indeed, 
as we will show this phase can exist when strong forward
scattering couplings between nearest and (at least)
second nearest neighbour chains are taken into account.

\paragraph*{The sliding Luttinger liquid.}

We consider an anisotropic 2D system composed of parallel chains
(labelled by an integer $i$) with spinless Luttinger liquids in
each chain. The bosonized form of the fermion operators near the
right and left fermi points are:  
\[
\psi_{\text{\tiny R/L},i}(x)= 
\frac{1}{\sqrt{4\pi \epsilon}}e^{i \phi_{\text{\tiny R/L},i }(x)}
\chi_{\text{\tiny R/L},i}
\]
 where $\epsilon$ is some intra-chain cut-off, the
$\chi_{\text{\tiny R/L},i}$ are the Klein factors and $x$ is the
coordinate along the chain. The Luttinger liquid on each chain
is described by the usual lagrangian:
\begin{equation}
\label{lzero}
{\mathcal L}_0^{(i)} =
\frac{K_0}{2}%\sum_{i=-\infty}^{\infty}
\left[(\partial_t
\Theta_i(x,t))^2 - (\partial_x \Theta_i(x,t))^2
\right]
\end{equation}
where $\Theta_i=(\phi_{L,i}-\phi_{R,i})/\sqrt{4\pi}$, and $K_{0}$ is an 
 interaction dependent constant with 
 $K_0>1$ for repulsive intra-chain interactions. 
The sound velocity on each chain has been set to unity. 
In the following, we will also use the dual field
$\Phi_i(x,t)=(\phi_{L,i}+\phi_{R,i})/\sqrt{4\pi}$.

We now add forward scattering interactions between the
chains which correspond to couplings between the long
wavelength components of the densities $\rho_i(x,t)$, 
and of the currents $J_i(x,t)$: 

\begin{equation}
\label{lint}
{\mathcal L}_{int} = 2\pi \sum_{i\ne j}[
J_i\bar{K}^J_{ij}J_j -
\rho_i\bar{K}^\rho_{ij}\rho_j]
\end{equation}
where the interactions are assumed local in $x$ and time. Using
the bosonization relations 
$\sqrt{4\pi}\rho_i(x,t)=\partial_x \Theta_i(x,t)$ 
and 
$\sqrt{4\pi}J_i(x,t)=-\partial_t\Theta_i(x,t)$, 
we obtain the bosonized form of the lagrangian 
${\mathcal L}_{tot} = \sum_{i}{\mathcal L}_0^{(i)}+{\mathcal L}_{int}$ 
for the interacting Luttinger liquids :
\begin{equation}\label{ltot}
{\mathcal L}_{tot} 
=\frac12
\sum_{i,j}\left[(\partial_t\Theta_i)K^J_{ij}(\partial_t\Theta_j) -
(\partial_x\Theta_i) K^\rho_{ij} (\partial_x\Theta_j)\right]
\end{equation}
where the coupling matrices are defined by 
$K_{ij}^J = K_0 \delta_{ij} + \bar{K}_{ij}^J,
K_{ij}^\rho = K_0 \delta_{ij} + \bar{K}_{ij}^\rho$. 
 By introducing Fourier transforms in the direction transverse to the
chains, this lagrangian can be rewritten as 
\begin{equation}
\label{ltot2}
{\mathcal L}_{tot} = \int_{q_\perp}
\frac{K({q_\perp})}{2}
\left[
\frac1{v(q_\perp)}|\partial_t\Theta_{q_\perp}|^2 - v(q_\perp)
|\partial_x\Theta_{q_\perp}|^2
\right]
\end{equation}
with the notation
$\int_{q_\perp}=\int^\pi_{-\pi}\frac{dq_\perp}{2\pi}$ (the transverse
chain spacing has been set to unity).
 The stiffness $K(q_\perp)$ is defined by 
$K(q_\perp)=\sqrt{K^J(q_\perp)K^\rho(q_\perp)}$ and the velocity 
$v(q_\perp)=\sqrt{K^\rho(q_\perp)/K^J(q_\perp)}$. Note that 
Lorentz invariance corresponds to   
$K^\rho(q_\perp)\propto K^J(q_\perp)$ (like in the isotropic
model studied in\cite{ohern99}). In that case,
 all modes have the same velocity. 

The lagrangian (\ref{ltot}) is invariant under the
transformations $\Phi_i \rightarrow \Phi_i + c_i$ and
$\Theta_i\rightarrow \Theta_i + d_i$ where $c_i$ and $d_i$ are
constants on each chain. We will thus call the
corresponding fixed point a 
{\emph{sliding Luttinger liquid (SLL)}} fixed point \cite{sliding}. 
 From this symmetry we deduce that the total numbers of
left (right) moving fermions on each chain are good quantum
numbers and expectation values of operators that change these -
such as $\langle \psi^\dagger_{L,i}\psi_{L,j} \rangle$ for $i \ne
j$ - are necessarily zero in this phase. This corresponds to a perfect
charge insulator in the transverse direction. Density (and current)
correlations in the transverse direction are however nontrivial. For
short ranged density and current interactions, they decay
exponentially with separation between the chains. The low energy modes
are density oscillations (sound) with dispersion
$E(q_\parallel,q_\perp)=v(q_\perp) |q_\parallel|$ (where $q_\parallel$
is the wavevector along the chain) which in general will propagate
both parallel and perpendicular to the chains. These modes can, for
instance, transport heat perpendicular to the chains although the
system is a perfect charge insulator in that direction. Correlation
functions along the chains exhibit power law behaviour as in the
Luttinger liquid. However the exponents now depend on the function
$K(q_\perp)$ rather than on a single number as in the case of
completely decoupled Luttinger liquids (\ref{lzero}). Therefore,
relations between exponents that are valid for decoupled Luttinger
liquids no longer hold in this case.

 As this phase is described by a gaussian lagrangian, we can study 
perturbatively the possible relevance of various operators to ascertain 
its stability. 

\paragraph*{Transverse hopping operators.}

 In the usual case of two chains, it is easily shown that the most
relevant operators correspond either to single particle (SP),
particle-hole (CDW) or pair hopping (SC) \cite{yakovenko92}. It is
thus natural to first focus on these operators in our stability
analysis. They are defined respectively by
\begin{subequations}
\label{hopop}
\begin{align}
\delta {\mathcal L}_{\text{\tiny SP}}&=\sum_{i  ,j } 
~t_{\perp}^{i  j } 
\left( 
 \psiRd{,i  }\psiR{,j } 
+\psiLd{,i  }\psiL{,j }
+ h.c. \right)
\\
\delta{\mathcal L}_{\text{\tiny CDW}}&= \sum_{i  j } 
~g^{i  j }_{\text{\tiny CDW}}
\left(
\psiRd{,i  }\psiL{,i  }
\psiLd{,j  }\psiR{,j  } 
+ h.c. \right)
\\
\delta {\mathcal L}_{\text{\tiny SC}}&= \sum_{i  j } 
~g^{i  j }_{\text{\tiny SC}}
\left(
\psiLd{,i  }\psiRd{,i }
\psiR{,j  }\psiL{,j  } 
+ h.c. \right)
\end{align}
\end{subequations}

 Upon bosonization, the corresponding operators read: 
\begin{subequations}\label{hopping-boson}
\begin{eqnarray}
{\mathcal O}_{\text{\tiny SP}}^{i  j } &=&
\cos  \sqrt{\pi}(\Phi_{i  }-\Phi_{j })
\cos \sqrt{\pi}(\Theta_{i  }-\Theta _{j })
\\
{\mathcal O}_{\text{\tiny CDW}}^{i  j }&=&
\cos \sqrt{4\pi}\left(\Theta_{i  }-\Theta_{j } \right)
\\
{\mathcal O}_{\text{\tiny SC}}^{i  j }&=&
\cos \sqrt{4\pi}\left(\Phi_{i  }-\Phi_{j } \right)
\end{eqnarray}
\end{subequations}

The dimensions of these operators at the SLL fixed point (\ref{ltot2})
 are readily evaluated and are given by : 
\begin{subequations}\label{dimensions}
\begin{align}
\label{eta-cdw}
\eta_{\text{\tiny CDW}}^{(N)}&
= 2 \int_{q_{\perp}} \frac{1-\cos(Nq_{\perp})}
{K(q_{\perp})} \\
\label{eta-sc}
\eta_{\text{\tiny SC}}^{(N)} &
= 2 \int_{q_{\perp}} (1-\cos(Nq_{\perp}))K(q_{\perp})\\
\eta_{\text{\tiny SP}}^{(N)} &=
\frac{1}{4}\left(\eta_{\text{\tiny CDW}}+\eta_{\text{\tiny SC}} \right)
\end{align}
\end{subequations}
 where $N=|i  -j|$. 
 A necessary condition for this sliding LL phase to be stable is
thus that 
\begin{equation}\label{stability}
\eta_{\text{\tiny CDW}}^{(N)},\eta_{\text{\tiny SC}}^{(N)}>2 \quad
\text{and} \quad \eta_{\text{\tiny CDW}}^{(N)}+\eta_{\text{\tiny SC}}^{(N)}>8 
\end{equation}
 for all $N \ge 1$.

\paragraph*{Stability of the SLL phase.} 
 Before discussing the stability of SLL fixed points, let us first
discuss the domain of validity of our RG approach. We are considering
the scaling dimension of perturbing operators around the fixed points
defined by the action (\ref{ltot2}).  This action is defined provided
the stiffness $K(q_{\perp})$ is positive everywhere in $[-\pi,\pi ]$.
Our study is thus restricted to the corresponding subspace of
 $K(q_\perp)$. At the boundary of this subspace $K(q_{\perp})$
vanishes for some $q_{\perp}^0\in [0,\pi]$ \cite{doubleroot}. The 
density correlations of transverse wavevector $q_{\perp}^0$ then diverge,
 signalling an instability towards transverse charge density wave ordering.
 In the transverse CDW, $\langle \rho_{q_{\perp}^{0}}\rangle \neq 0$, so then 
the total charge density on each chain is a function of the transverse 
position $i$ \cite{XXZ}. That the boundary corresponding to this transition 
will play a crucial role in  the
following analysis can be seen by inspecting the dimensions
(\ref{dimensions}) : $\eta_{\text{\tiny CDW}}^{(N)}$ can be
significantly increased by the presence of the pole in the integrand 
(root of $K(q_{\perp})$) with $\eta_{\text{\tiny SC}}^{(N)}$ being not
much affected \cite{ohern99}. Hence it may be possible to have these operators
irrelevant (i.e. with dimension greater then $2$),
for parameters in the vicinity of this boundary.

\begin{figure}[thb]
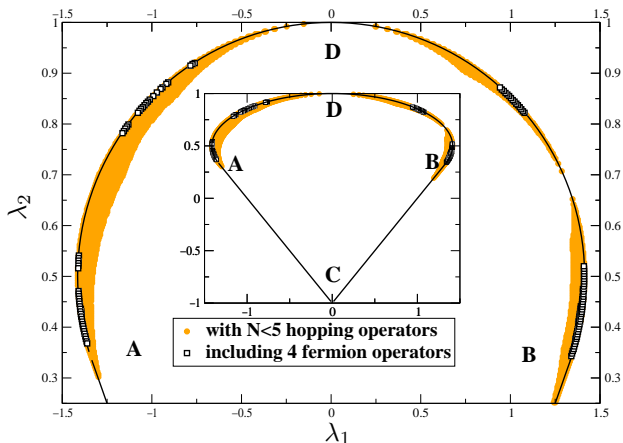

\psfrag{k1}{$\lambda_{1}$}
\psfrag{k2}{$\lambda_{2}$}
\centerline{\fig{8cm}{figure1.eps}}
\caption{SLL fixed points for the model in (\ref{modelK}). The allowed
parameter range for which $K(q_{\perp})$ is positive is within the
curve ACBD. For each point shown, there exists a finite range of $K_0$
for which the SLL is found to be stable against (a) CDW, SC and SP
perturbing operators of eq.(\ref{hopop}) with $N\le 4$ shown by the
small dots and (b) including general four point operators with $N\le
10$ shown as squares.
}
\label{fig1}
\end{figure}

{\it Stability analysis of a model $K(q_{\perp})$}
Here we shall consider the stability analysis for a concrete
model of $K(q_{\perp})$. 
 We thus look for a natural restriction to a finite number of terms of
the  Fourier expansion, that may allow a stable SLL fixed point. The
simplest case turns out to be 
\begin{equation}
\label{modelK}
K(q_{\perp})= 
K_{0}[1+\lambda_{1}\cos (q_{\perp })+\lambda_{2}\cos (2q_{\perp
})]
\end{equation}
(As we will discuss later, the model with just $\lambda_1$ included
does not possess a stable SLL phase). The requirement of positivity of
$K(q_{\perp})$ restricts the range of $\lambda_1$ and
$\lambda_2$, as shown in the Figure \ref{fig1}, to the region within
the boundary ADBC (henceforth simply denoted by ${\mathcal B}$). 
The stability of the SLL to the perturbations 
(\ref{hopping-boson}) with $1\le(N=|i -j|)\le4$ is first
determined \cite{operator-set}. The exponents in eq. (\ref{dimensions})
are numerically evaluated and the results are shown in Figure
\ref{fig1} which is a two dimensional representation of the
$(K_0,\lambda_1,\lambda_2)$ space. The points marked are those for which
there exists a range of $K_0$ values where the above operators are all
irrelevant. The corresponding values of $K_{0}$ are all greater than
one (repulsive interactions on the chains). 
 All the stable fixed points are thus found close to the boundary
$\mathcal{B}$.

 In an attempt to define stable RG fixed points, one may worry
about more general perturbing operators.  
Indeed, inclusion of the operators (\ref{hopping-boson}) 
with $5\leq N\leq 10$ does not significantly change the results. 
However, it turns out that general four point 
operators such as, for example, simultaneous hopping of 
fermions from chain $i  $ to $i  +1$ and from
$j$ to $j +1$ \cite{fourpoint}, has a dramatic effect on
the stability region  which is now much reduced. 
Thus, these operators are found to be relevant at a 
large number of the previously stable points. 
There are however some remaining fixed points, 
clustered close to the boundary ${\mathcal B}$, which
are shown as squares in Figure \ref{fig1}.

\begin{figure}[thb]
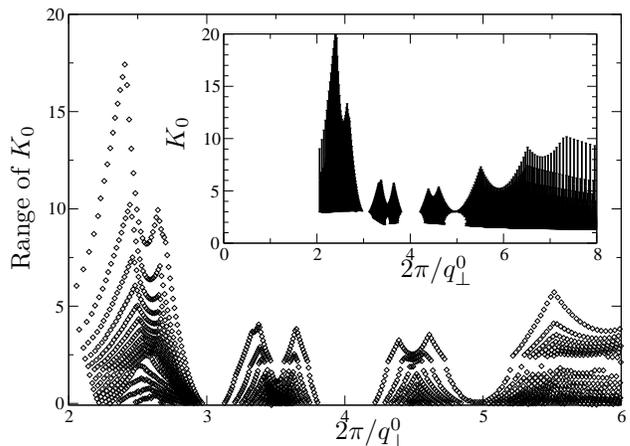

\psfrag{xxx}{$2\pi / q_{\perp}^{0}$}
\psfrag{yyyyyyyyy}{\hspace{-0.4 cm}Range of $K_{0}$}
\psfrag{zzzz}{$K_{0}$}
\centerline{\fig{8cm}{figure2.eps}}
\caption{Range of $K_{0}$ corresponding to stable SLL fixed points as
a function of the transverse CDW wavelength $2\pi /
q_{\perp}^{0}$. Inset shows actual values of $K_{0}$
}
\label{fig2}
\end{figure}
 
 Thus, as expected from the argument given above, all stable fixed
points are  found close to the boundary ${\mathcal B}$. This can be
physically understood as follows. 

 Let us recall that anywhere on  ${\mathcal B}$ the system is on the
verge of a transverse CDW instability.  
Such a transverse CDW would frustrate crystallization of the fermions since 
$(2k_F)^{-1}$, which controls the spacing of the particles on each
chain,  is now a function of the chain index $i$. 
Strong fluctuations of this kind prevent the locking in of density
modulations along the chain, 
hence defeating the crystal instability and stabilizing the SLL. 
This mechanism will be less effective if the wavevector $q^0_\perp$ is
commensurate with the transverse lattice {\it i.e.} if $q_{\perp}^0=2\pi
m/n$ for some integers $m$,$n$. Then, the longitudinal density
waves on chains separated by $n$  can lock in and give rise
to the crystal instability (provided $n$ is not too large).  
This idea is confirmed on inspecting the range of $K_{0}$ for 
which the SLL phases exist. 
In Figure \ref{fig2} the corresponding range of $K_0$ 
is plotted as a function of the transverse CDW instability period 
$2\pi/q_\perp^0$ (that it is nearest to). Drastic reduction of the SLL
stability occurs at 
commensurate transverse wavevectors. This also allows us to
understand the absence of stable SLL in the model with
$\lambda_2=0$. Then, the only existing transverse CDW has 
wavevector $\pi$ (for $\lambda_1=1$) and hence for strong enough repulsive
 interactions, we expect the longitudinal density waves on next
nearest neighbor chains (N=2) to lock and lead to a crystalline
instability. This turns out to be a correct expectation as, by
including the N=2 CDW operator, we do not find a stable SLL in this
model.

     The physical mechanism identified here allows us to generalize
our results to models beyond the simple ones considered so far,
including
 Luttinger liquids coupled in three dimensions. In general 
we expect the SLL to be stabilized for renormalized
couplings in the vicinity of an incommensurate transverse CDW
transition. 

 An approach very similar to the one taken in this paper can be applied
to bosonic models at any filling, coupled anisotropic spin chains, 
or vortex lattices in anisotropic
superconductors. These, together with 
the more complicated case of spinful fermions will be discussed in a
forthcoming publication\cite{inprep}. 
 As disorder in known to modify significantly the behaviour of a
single Luttinger liquid, it may also be highly interesting to study its
effect on the new phase described in this paper. 

In conclusion we have studied in detail the occurrence of the
sliding Luttinger liquid phase in a simple three parameter model.
We find that the region of stability of this phase is restricted
to be close to the boundary where the harmonic boson theory
breaks down and a soft mode appears, that signals a transverse CDW
instability. Strong incommensurate transverse CDWs are
particularly effective at frustrating crystal formation and
enhancing the stability of the SLL. We expect these conclusions
to apply also to more general models than the simple one
considered in this paper.

{\it During the completion of this work, we became aware of a
similar, but nevertheless different, study by Emery et al. \cite{emery00}. }

 It is a pleasure to thank T. Senthil for suggestions on this
problem and continuous moral support, and L. Balents, M.P.A.
Fisher, K. Damle, A. Lopatin, T. Giamarchi, S.L. Sondhi and
F.D.M. Haldane  for discussions, and I. Gruzberg for a carefull
reading of the manuscript. 
One of us (A.V.) would like to thank the ITP, Santa Barbara for
hospitality during part of this work. This work was supported by the
NSF grant No. DMR-9528578.

%%%%%%%%%%%%%%%%%%%%%%%%%%%%%%%%%%%%%%%%%%%%%%%%%%%%%%%%%%%%%%%%%
%%%%%%%%%%
%%%%%%%%%%%%%%%%%%%%%%%%%%%%%%%%%%%%%%%%%%%%%%%%%%%%%%%%%%%%%%%%%
%%%%%%%%%%
%%%%%%%%%%%%%%%%%%%%%%%%%%%%%%%%%%%%%%%%%%%%%%%%%%%%%%%%%%%%%%%%%
%%%%%%%%%%
%%%%%%%%%%%%%%%%%%%%%%%%%%%%%%%%%%%%%%%%%%%%%%%%%%%%%%%%%%%%%%%%%
%%%%%%%%%%
%%%%%%%%%%%%%%%%%%%%%%%%%%%%%%%%%%%%%%%%%%%%%%%%%%%%%%%%%%%%%%%%%
%%%%%%%%%%
%%%%%%%%%%%%%%%%%%%%%%%%%%%%%%%%%%%%%%%%%%%%%%%%%%%%%%%%%%%%%%%%%
%%%%%%%%%% 

\unecol 

\begin{thebibliography}{99}
\bibitem{QHEstripefootnote} Although the electrons in QHE stripes are
spin polarized, the physics in that case differs in many details from
what we consider here, due to the presence of the soft translational
mode and the magnetic field.
\bibitem{sliding} Thus in the sliding Luttinger liquid, both the
CDW phase field $\Theta_i$ and the Josephson phase field $\Phi_i$
can slide relative to each other at no cost.
\bibitem{doubleroot} The case of double roots is considered in 
\cite{inprep}.
\bibitem{operator-set} Choosing the set of perturbing operators
for a physical situation is an involved issue, which will depend
on details of the system and the small but nevertheless finite
temperature at which it is considered. Then, operators that have
small bare values and are not further generated in the RG could
be unimportant, even if relevant.
\bibitem{fourpoint} More precisely, we consider operators of the
form $\psi^\dag_{\alpha   i}\psi^\dag_{\beta j}\psi_{\gamma
k}\psi_{\delta l}$ where $\alpha  =\pm1$ for (L/R) and
$\alpha  +\beta=\gamma+\delta$ by momentum conservation. We have
taken the maximum separation between any two fermion operators to
be $\le 10$. The ones that destabilize the SLL points usually
involve simultaneous hops on pairs of nearest neighbour chains.
These could well be present or generated in physical systems.
\end{thebibliography}
\end{document}